# Self-consistent calculation of semiconductor heterojunctions by using quantum genetic algorithm

**Mehmet ŞAHİN[1*], Mehmet TOMAK[2]**

[1]Selcuk University, Faculty of Arts and Science, Dept. of Physics Kampus, 42075 Konya, Turkey

[2]Middle East Technical University, Dept. of Physics Ankara, 06531 Turkey

**Abstract**

In this study, we have investigated the ground state energy level of electrons in modulation doped $GaAs/Al_xGa_{1-x}As$ heterojunctions. For this purpose, Schrödinger and Poisson equations are solved self consistently using quantum genetic algorithm (QGA). Thus, we have found the potential profile, the ground state subband energy and their corresponding envelope functions, Fermi level, and the amount of tunneling charge from barrier to channel region. Their dependence on various device parameters are also examined.

*Key words:* heterostructures, 2DEG, genetic algorithm, self-consistent

*PACS:* 02.60.Pn, 02.90.+p, 73.20.Dx, 73.40.Kp

## 1. Introduction

The two-dimensional electron gas (2DEG) of a modulation-doped $GaAs/Ga_{1-x}Al_xAs$ heterostructure is readily formed in electronic devices such as high electron mobility transistor (HEMT) and quantum well infrared photodetector (QWIP) [1-3]. Studies of the energy levels, electron mobility and optical properties of 2DEG using analytical and numerical approaches have been reported in numerous studies [4-13]. Stern [14] and Ando [15] have solved 2DEG problem numerically within the Hartree and density functional approximations. Bastard [16] has applied the variational self-consistent method for the electric quantum limit (EQL) in the Hartree approximation using generalized Fang-Howard [17] wave functions. Since then, many electronic structure calculations of 2DEG have been performed using different methods [6,18-21]. The pioneering works on variational calculations of 2DEG were based on various approximations such as the neglect of the tunneling to the barrier region [4], the employment of the finite or infinite triangular potential wells [6] and the Hartree approximations [16]. The excellent reviews were presented by Ando et.al [22], Hiyamizu [23] and Weisbuch [24].

---

[*] e-mail address: sahinm@selcuk.edu.tr

In this study, we present self-consistent calculations of ground state energy level of 2DEG with quantum genetic algorithm (QGA) which is based on energy minimization. QGA method has been used in many different fields such as nonlinear fitting problem [25], crystal growth [26], quantum mechanical systems with one and two particles [27-29], and atomic physics [30]. So far, however, to the best of our knowledge QGA method has not been applied to any self-consistent heterojunction problems. This aim of this study is to investigate the applicability of the QGA-method to a complicated, realistic self-consistent heterojunction problem.

This paper is organized as follows: The next section, presents a brief theory and formulation. In section 3, description of QGA is presented. Results and discussion is given in the last section.

## 2. Theory and Formulation

As well known, to determine the energy levels and charge transfer in a single heterojunction, coupled Poisson and Schrödinger equations have to be solved with self-consistently,

$$\left[ -\frac{\hbar^2}{2} \frac{\partial}{\partial z} \frac{1}{m(z)} \frac{\partial}{\partial z} + V_b(z) - eV_{sc}(z) \right] \psi_i(z) = E_i \psi_i(z), \qquad (1)$$

$$\frac{d^2 V_H(z)}{dz^2} = \frac{4\pi e}{\kappa(z)} \left[ \sum_i n_i \psi_i^2(z) - N_D^+ + N_A^- \right], \qquad (2)$$

where

$$V_{sc} = V_H(z) + V_{xc}(z), \qquad (3)$$

and $V_H(z)$ is the Hartree potential, $V_{xc}(z)$ is the exchange-correlation potential, $V_b(z)$ is the barrier potential, $n_i$ is the areal concentration of electrons in the *i*th subband, $N_D^+$ and $N_A^-$ are the donor and acceptor concentration respectively, $\kappa(z)$ and $m(z)$ are the position-dependent static dielectric constant and effective mass respectively.

At finite temperature T, the chemical potential (or Fermi level) of the electrons μ (or $E_F$) and the quantities $n_i$, $E_i$, $m_i$ are related by:

$$n_i = m_i \frac{k_B T}{\pi \hbar^2} \ln\left[1 + \exp\left(\frac{\mu - E_i}{k_B T}\right)\right], \qquad (4)$$

where $k_B$ is the Boltzmann constant. At T=0 K, this equation reduces to

$$n_i = \frac{m_i}{\pi \hbar^2}(\mu - E_i)\Theta(\mu - E_i), \qquad (5)$$

where $\Theta$ is the step function. The above set of equations needs to be completed by boundary conditions. The bound state envelope function $\psi_i(z)$ should go to zero while $z \to \pm\infty$ and $\frac{1}{m(z)}\frac{\partial \psi_i}{\partial z}$ should be continuous everywhere. As for the Poisson equation, it is required that the heterojunction be in electrical equilibrium, namely

$$\sum_i n_i + \int_{-\infty}^{\infty}\left(N_A^-(z) - N_D^+(z)\right)dz = 0. \qquad (6)$$

In addition, the heterojunction is in thermodynamical equilibrium. This condition requires the chemical potential to be constant.

At T=0 K, the sharpness of the Fermi-Dirac distribution function provide important simplification in calculation of the contributions of donors and acceptors to the Hartree potential. So, the contributions due to donors and acceptors are

$$V_A(z) = \frac{2\pi e}{\kappa} N_A^- z(z - 2\ell_A), \qquad (7)$$

$$V_D(z) = -\frac{2\pi e}{\kappa}(N_D - N_{b,A})(z + \ell_D + w)^2 + V_0, \qquad (8)$$

where $\ell_A$ and $\ell_D$ are acceptor and donor depletion lengths respectively, w is the spacer layer thickness, $N_{b,A}$ is the residual acceptor concentration in the barrier region and $V_0$ integration constant which has to be determined by continuity of the electrostatic potential at z=−w. For GaAs channels containing $10^{14}$cm$^{-3}$ acceptors, one finds $\ell_A \cong 4.6\mu m$ [32]. The spatial extension of the $\psi_i$'s is quite smaller than this length and therefore, in the solution of the Schrödinger equation, $V_A(z)$ may safely be approximated by the linear relation

$$V_A(z) = -\frac{4\pi e}{\kappa} N_{dep} z, \qquad (9)$$

where $N_{dep} = N_A^- \ell_A$. In the depletion length approximation, the charge balance equation can be written as

$$N_D \ell_D = \sum_i n_i + N_{dep} + N_{b,A}(w + \ell_D). \qquad (10)$$

The transferred charge is $N_s=\sum_i n_i$. The donor depletion length can be determined self-consistently from

$$\mu = V_b - E_D - \frac{4\pi e^2}{\kappa} n_i \int_{-\infty}^{0} z\psi_i^2(z)dz - \frac{2\pi e^2}{\kappa}\{(N_D - N_{b,A})\ell_D(\ell_D + 2w) - N_{b,A}w^2\}, \quad (11)$$

where $E_D$ is donor binding energy. Here, we assume that $N_{b,A}=0$.

The potential arising from electron-electron interaction is calculated by finite difference iteration method. In this method, Poisson equation can be given as

$$\frac{d^2 V_{el-el}^j}{dz^2} = \frac{V_{el-el}^{j-1} - 2V_{el-el}^j + V_{el-el}^{j+1}}{h^2} = \rho_j(z), \quad (12)$$

where

$$\rho_j(z) = \frac{4\pi e}{\kappa} \sum_i n_i \psi_i^2(z). \quad (13)$$

Here, the index j is the mesh point and h is distance at between two adjacent mesh points.

We follow QGA to solve the coupled Poisson and Schrödinger (PS) equations self-consistently.

## 3. Genetic Process and Calculations

Genetic algorithms (GA) are general search and numerical optimization algorithms inspired by both natural selection and genetics. This approach is gaining a growing attraction in the physical and computer sciences and in engineering. GA, is firstly proposed by Holland [31]. Recently, this method has appeared to be used more frequently in the optimization and minimization problems for the quantum mechanical systems [27-29]. GA process is established on three basic principles; reproduction (or copy), crossover and mutation.

In this study, we use modified Fang-Howard [17,32] trial wavefunction, which allows penetration into the barrier region and obeys the boundary condition mentioned in section 2. This wavefunction is

$$\psi_k(z) = \begin{cases} A\exp(\alpha_k z/2) & z \leq 0 \\ B(z+z0)\exp(-\beta_k z/2) & z \geq 0 \end{cases}, \quad (14)$$

with random values for $\alpha_k>0$ and $\beta_k>0$. A and B are normalization constants and they determined numerically from

$$\int_{-\infty}^{\infty} \psi_k^*(z)\psi_k(z)dz = 1. \quad (15)$$

We have chosen population number (npop) as 100. Initial population has been created numerically from Eq. (14) for random values of $\alpha_k$ and $\beta_k$ (k=1…npop) and assigned to two dimensional vector arrays. This population has been normalized by using Eq. (15). Thus, a normalized random population of wavefunctions (individuals) is created as an initial generation. Expectation value of energy is determined from this generation by means of

$$E_k = \langle \psi_k(z) | \hat{H} | \psi_k(z) \rangle. \tag{16}$$

Fitness values are created by using these energy values. For this aim, we use the following expression.

$$\text{Fitness}[\psi_k] = \exp[-\sigma(E_k - E_{av})], \tag{17}$$

where, $\sigma$ is a constant and $E_{av}$ is the average of the energy eigenvalues. By using this fitness values, a rulet wheel [33] is constituted and a selection procedure has been performed. In this selection procedure, usually, better individuals are selected, however sometimes less fit individuals can also be selected and new generation is created from this set of chosen individuals. This process is usually known as reproduction or copy.

In the crossover, we take two randomly chosen individuals (or wavefunctions). Two new functions are produced by using these individuals as

$$\psi'_1(z) = \psi_1(z)\text{cr}(z) + \psi_2(z)[1 - \text{cr}(z)]$$

$$\psi'_2(z) = \psi_2(z)\text{cr}(z) + \psi_1(z)[1 - \text{cr}(z)], \tag{18}$$

where cr(z) is a smooth step function [27,34] or its value can be randomly selected from a uniform distribution between (0,1). At each step of GA iteration, kind of crossover (namely the selection of the smooth step function or a random real number) is randomly determined. In our problem, we have seen that this crossover scheme is more efficient. For this problem the probability of crossover type has been chosen as 0,50 and the probability of crossover has been chosen in $0.10 \leq P_c \leq 0.20$. If the crossover probability is chosen larger than 0.20, the system can result in inappropriate solutions.

In mutation operation, random values were assigned to $\alpha$ and $\beta$ parameters at Eq. (14). In this way, $\psi_M(z)$ mutation function was constituted and added this function to any other randomly chosen $\psi_k(z)$ function to create a new parent function as

$$\psi'_k(z) = \psi_k(z) + C\psi_M(z), \tag{19}$$

where C is an amplitude of mutation function. We have selected the mutation probability as small as, $P_m \leq 0.005$.

During all of the GA iteration, copy or reproduction, crossover and mutation operations were randomly performed over the individuals. After the application of the genetic operations, new populations obtained were normalized.

In this study, all calculations are performed numerically. The derivatives are calculated over the five mesh points. In computation of the integrals, Simpson's method was used. For the potential calculation due to electron-electron interaction, the wavefunction chosen corresponds to the best fitness. However, the best fitness, has not been carried to new generation because of the nature of self-consistent calculation. Our algorithm may be summarized briefly as following: i) Firstly, initial population is created and normalized. ii) The expectation values of energy are determined for each individual in the barrier potential. Hartree potential initially is taken as zero. iii) Fitness values are computed with these energy eigenvalues and the best fitness determined. iv) A new generation is created from old one with genetic operations (copy or reproduction, crossover and mutation operations) and then normalized. v) Poisson's equation is solved using the wavefunction which corresponds to the best fitness vi) Hartree potential calculated is added to the barrier potential and returned back to step ii. This process is repeated until the best convergence is obtained.

## 4. Results and Discussion

The parameters used in this study are listed in Table 1. These parameters are taken from experimental work of Hiyamizu et. al. [35]. We have used atomic units at all calculations, where $\hbar=1$, the electronic charge e=1 and the electron mass m=1. We have represented the effective masses of electrons inside GaAs and AlGaAs as $m_1$ and $m_2$, and similarly dielectric constants as $\kappa_1$ and $\kappa_2$, respectively.

$$m^*(z) = \begin{cases} 1, & z > 0 \\ \dfrac{m_2}{m_1}, & z < 0 \end{cases} \quad , \quad \kappa(z) = \begin{cases} 1, & z > 0 \\ \dfrac{\kappa_2}{\kappa_1}, & z < 0 \end{cases}. \quad (20)$$

In Figure 1, the calculated self-consistent potential $V_{sc}$, ground state subband energy and Fermi energy level are shown. In the inset, tunelling electron concentrations $N_s$, determined self-consistently is shown. In this calculation, the exchange and correlation energies are not considered. Also the ground state subband wavefunction, which is determined by QGA, is given in this figure.

In Figure 2, the evolution of energy eigenvalue with the number of iterations is plotted. As seen from this figure, the variation of the energy eigenvalue with iteration shows initially an oscillatory behavior, then it converges to a constant value after nearly 60 iterations. The reason for this oscillation is the fact that the best fitness is not carried to the new generation because of the self-consistent procedure.

Figure 3 shows the evolution of the concentration of 2DEG versus the number of iterations. Here the oscillatory behavior is similar to Figure 2. These oscillations are due to the electrical and thermodynamical nonequilibrium. After some 50 iterations, system reaches to the electrical equilibria.

Figure 4 shows the 2DEG concentrations, determined self-consistently, as a function of spacer layer thickness w. As seen from the figure, charge transfer decreases as the spacer thickness w increases. The experimental results are taken from Hiyamizu et al. [35]. The agreement between experimental and calculated values is rather good.

Figure 5 shows the dependence of self-consistently calculated $N_s$ on the barrier height $V_b$ for different spacer layer thicknesses. As seen from the figure, 2DEG concentration is increasing with $V_b$ for all spacer layer thicknesses. An increase at $V_b$ corresponds to enlarge the energy separation between the donor level in AlGaAs and the ground subband level in GaAs. So, more electrons are transferred to the GaAs region.

So far, quantum mechanical applications of GA are usually concentrated on single particle systems [27-29]. However, there have been some reported studies performed on the two particles systems [27,30]. This study, to the best of our knowledge, is the first application of this method to the 2DEG problem.

All results are in agreement with the literature [32,36] and experiment [35]. As shown in our results, QGA method is quite efficient for the self-consistent heterojunction problem. So, this method can be applied to the calculation of the electronic structure of quantum nanostructures.

We have neglected the exchange-correlation term and image term at the calculations. The effect of image term is apparently extremely small for this materials and is discussed in detail by Stern and Das Sarma [8].

**Acknowledgment**

This study is partially supported by Selcuk University under grant No. BAP2001/112.

**References**

[1] L. Thibaudeau, P. Bois, J. Y. Duboz, J. Appl. Phys. **79**, 446 (1996)

[2] H. C. Liu, Semiconductors and Semimetals **62**, 129 (2000)

[3] H. C. Liu, Physica E **8**, 170 (2000)

[4] J. A. Pals, Phys. Lett. A **39**, 101 (1972)

[5] F. Stern, Phys. Rev. B **5**, 4891 (1972)

[6] A. A. Grinberg, Phys. Rev. B **32**, 4028 (1985)

[7] J. A. Appelbaum, G.A. Baraff, Phys. Rev. B **4**, 1235 (1971)

[8] F. Stern, S. Das Sarma, Phys. Rev. B **30**, 840 (1984)

[9] W. Walukiewicz, H. E. Ruda, J. Lagowski, H. C. Gatos, Phys. Rev. B **30**, 4571 (1984)

[10] M. P. Stopa, S. Das Sarma, Phys. Rev. B **47**, 2122 (1993)

[11] L. Hsu, W. Walukiewicz, Phys. Rev. B **56**, 1520 (1997)

[12] A. M. C. Serra, H. A. Santos, J. Appl. Phys. **70**, 2734 (1991)

[13] B. Vinter, Appl. Phys. Lett. **44**, 307 (1984)

[14] F. Stern, Bull. Am. Phys. Soc. **28**, 447 (1983)

[15] T. Ando, J. Phys. Soc. Jpn. **51**, 3893 (1982)

[16] G. Bastard, Surface Science **142**, 284 (1984)

[17] F. F. Fang and W. E. Howard, Phys. Rev. Lett. **16**, 797 (1966)

[18] J. E. Hasbun, Phys. Rev. B **43**, 5147 (1991)

[19] S.L. Ban, X. X. Liang, J. Luminescence **94**, 417 (2001)

[20] F. Sacconi, A. Di Carlo, P. Lugli, H. Morkoc, Phys. Stat. Sol. A **188**, 251 (2001)

[21] B. Jogai, J. Appl. Phys. **91**, 3721 (2002)

[22] T. Ando, A.B. Fowler, F. Stern, Rev. Mod. Phys. **54**, 437 (1982)

[23] S. Hiyamizu, Semiconductors and Semimetals **30**, 53 (1990)

[24] C. Weisbuch, Semiconductors and Semimetals **24**, 1 (1987)

[25] A. Brunetti, Comp. Phys. Comm. **124**, 204 (2000)

[26] Ö. Şahin, P. Sayan, A. N. Bulutcu, J. Cryst. Growth **216**, 475 (2000)

[27] I. Grigorenko, M. E. Garcia, Physica A **284**, 131 (2000); Physica A **291**, 439 (2001)

[28] H. Nakanishi, M. Sugawara, Chem. Phys. Lett. **327**, 429 (2000)

[29] R. Saha, P. Chaudhury, S. P. Bhattacharyya, Phys. Lett. A **291**, 397 (2001)

[30] L. Liu, L. Zhao, Y. Mao, D. Yu, J. Xu, Y. Li, Int. J. Mod. Phys. C **11** (2000) 183

[31] J. H. Holland, Adaptation in Natural and Artifical Systems, University of Michigan Press, (1975)

[32] G. Bastard, Wave Mechanics Applied to Semiconductor Heterostructures (Les Editions de Physique, Paris, 1988)

[33] D. E. Goldberg, Genetic Algorithms in Search, Optimization, and Machine Learning, Addison-Wesley, Reading, MA, (1999)

[34] H. Şafak, M. Şahin, B. Gülveren, M. Tomak, submitted for publication

[35] S. Hiyamizu, J. Saito, K. Nanbu, T. Ishikawa, Jpn. J. Appl. Phys. **22**, L609 (1983); S. Hiyamizu, T. Mimura, T. Fujii, K. Nanbu, H. Hashimoto, Jpn. J. Appl. Phys. **20**, L245 (1981)

[36] E. F. Schubert, K. Ploog, IEEE Trans. Elec. Dev. **ED-32**, 1868 (1985)

**Table 1.** Parameters used in calculations

| | |
|---|---|
| Al content | x=0.3 |
| Acceptor concentration in GaAs | $N_{dep}$=5.0.10$^{10}$ cm$^{-2}$ |
| Donor concentration in AlGaAs | $N_D$=2.0.10$^{18}$ cm$^{-3}$ |
| Spacer layer thickness | w=60 Å |
| Donor binding energy | $E_D$=60 meV |
| Effective masses | $m_1$=m(GaAs)=0.070$m_0$ |
| | $m_2$=m(AlGaAs)=0.088$m_0$ |
| Dielectric constants | $\kappa_1$=κ(GaAs)=13.1 |
| | $\kappa_2$=κ(AlGaAs)=12.2 |
| Barrier height | $V_b$=225 meV |

**Figure Captions**

**Figure 1.** The calculated self-consistent potential, the ground subband, Fermi energy and their correspond envelope function.

**Figure 2.** The evolution of energy eigenvalue with the number of iterations.

**Figure 3.** The evolution of 2DEG concentration with the number of iterations.

**Figure 4.** The calculated 2DEG concentration as a function of spacer layer thickness.

**Figure 5.** The dependence of concentration of 2DEG on the barrier height for different spacer layer thickness.

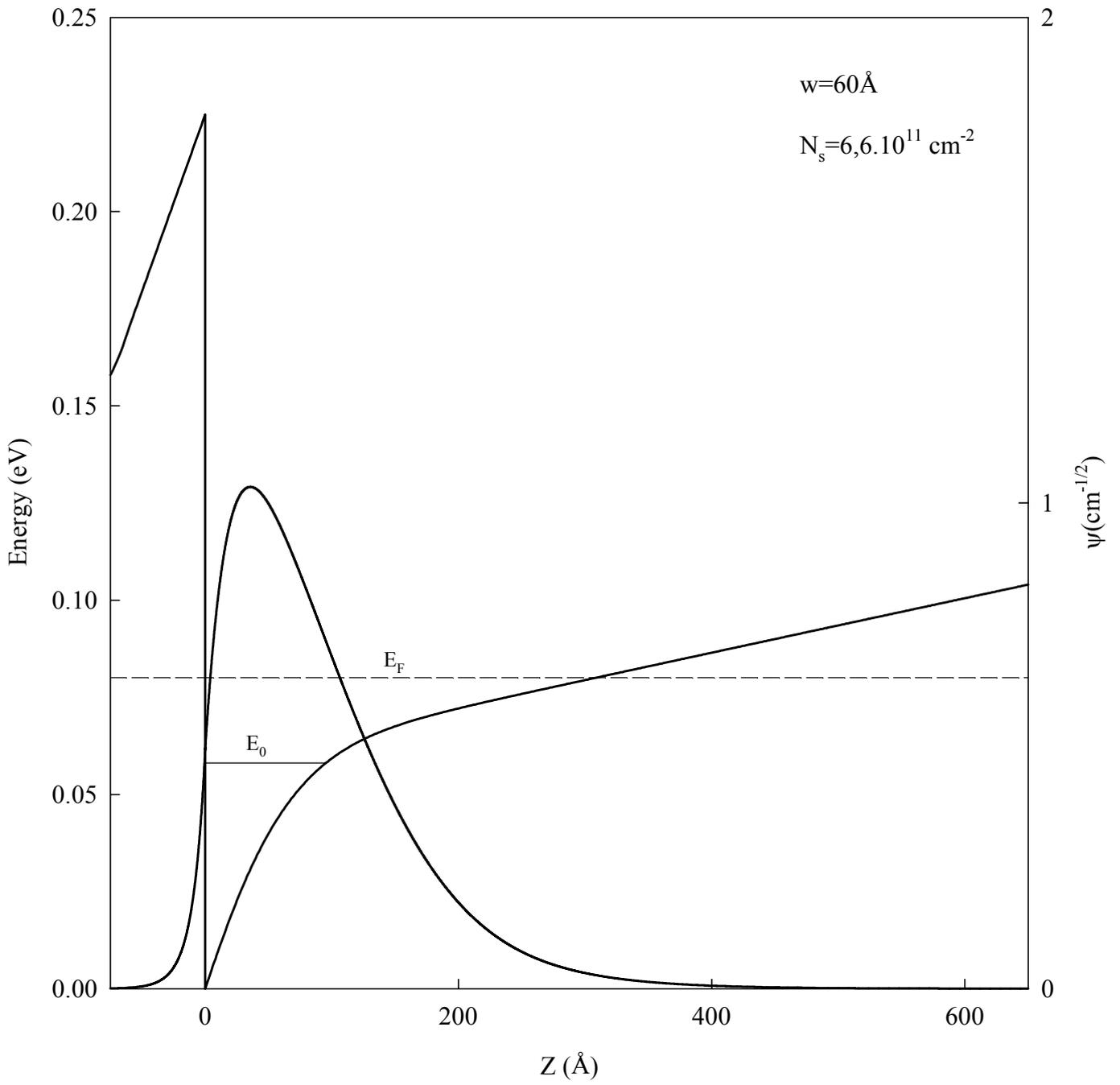

**Figure 1**

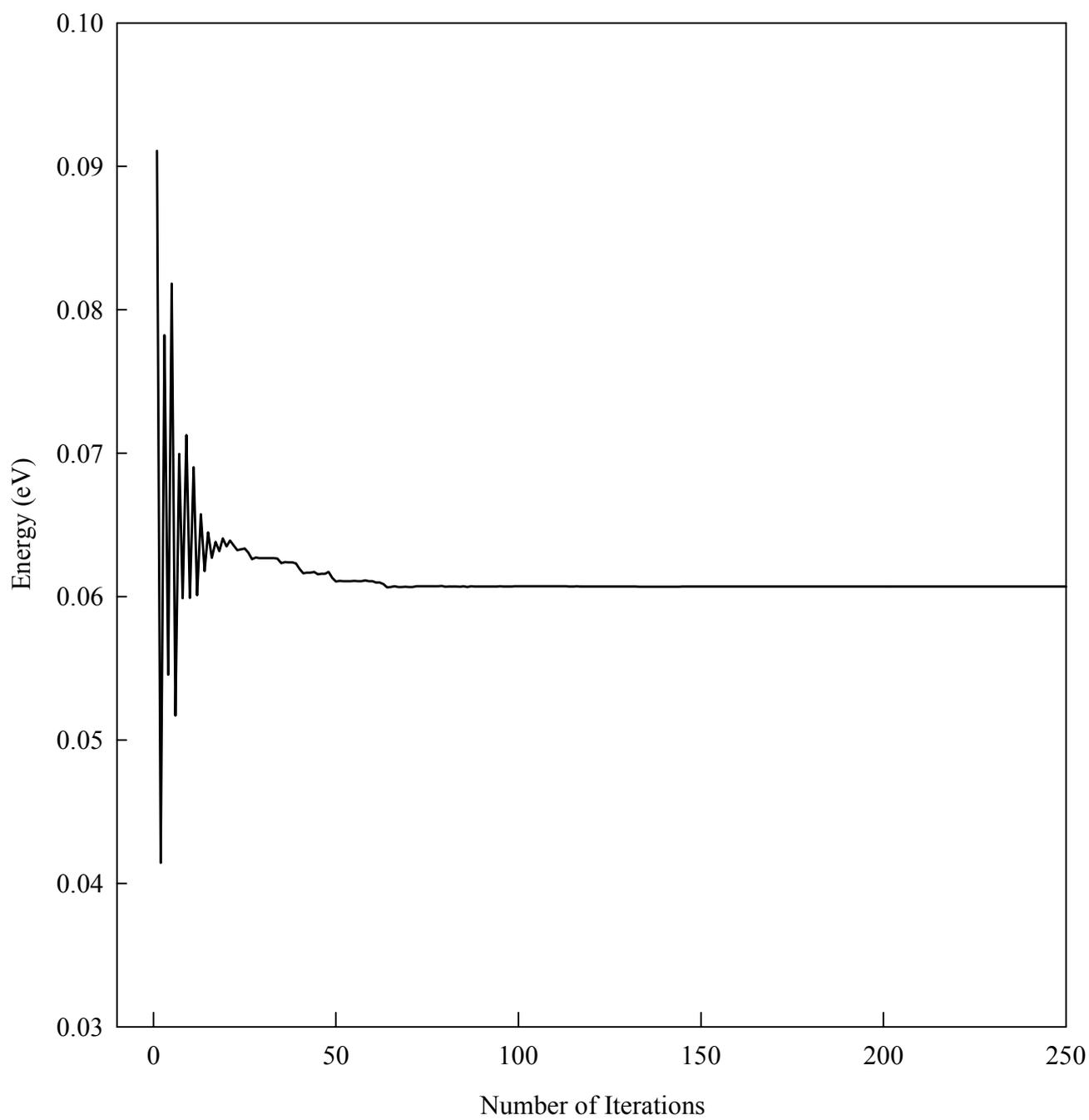

**Figure 2**

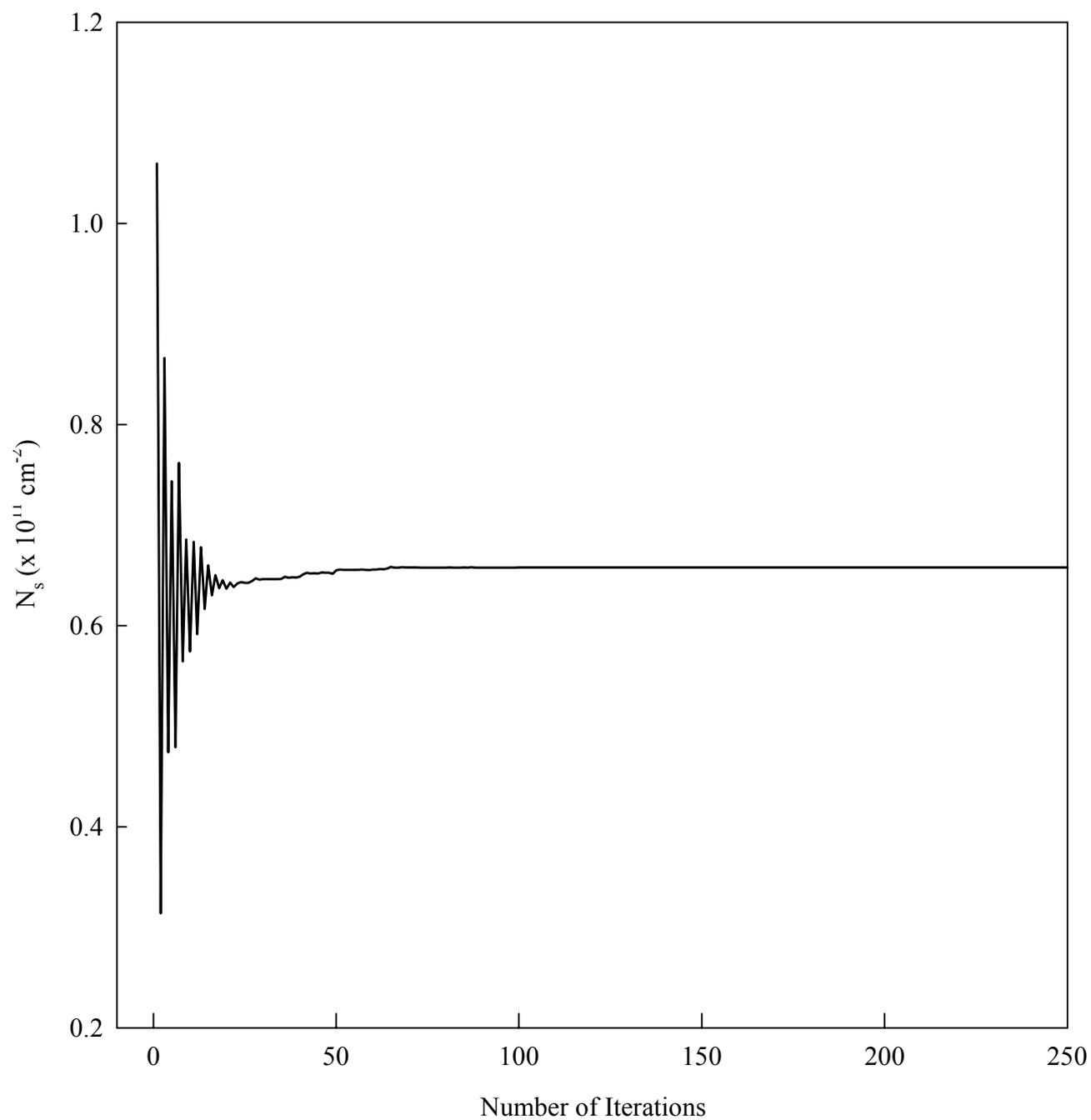

**Figure 3**

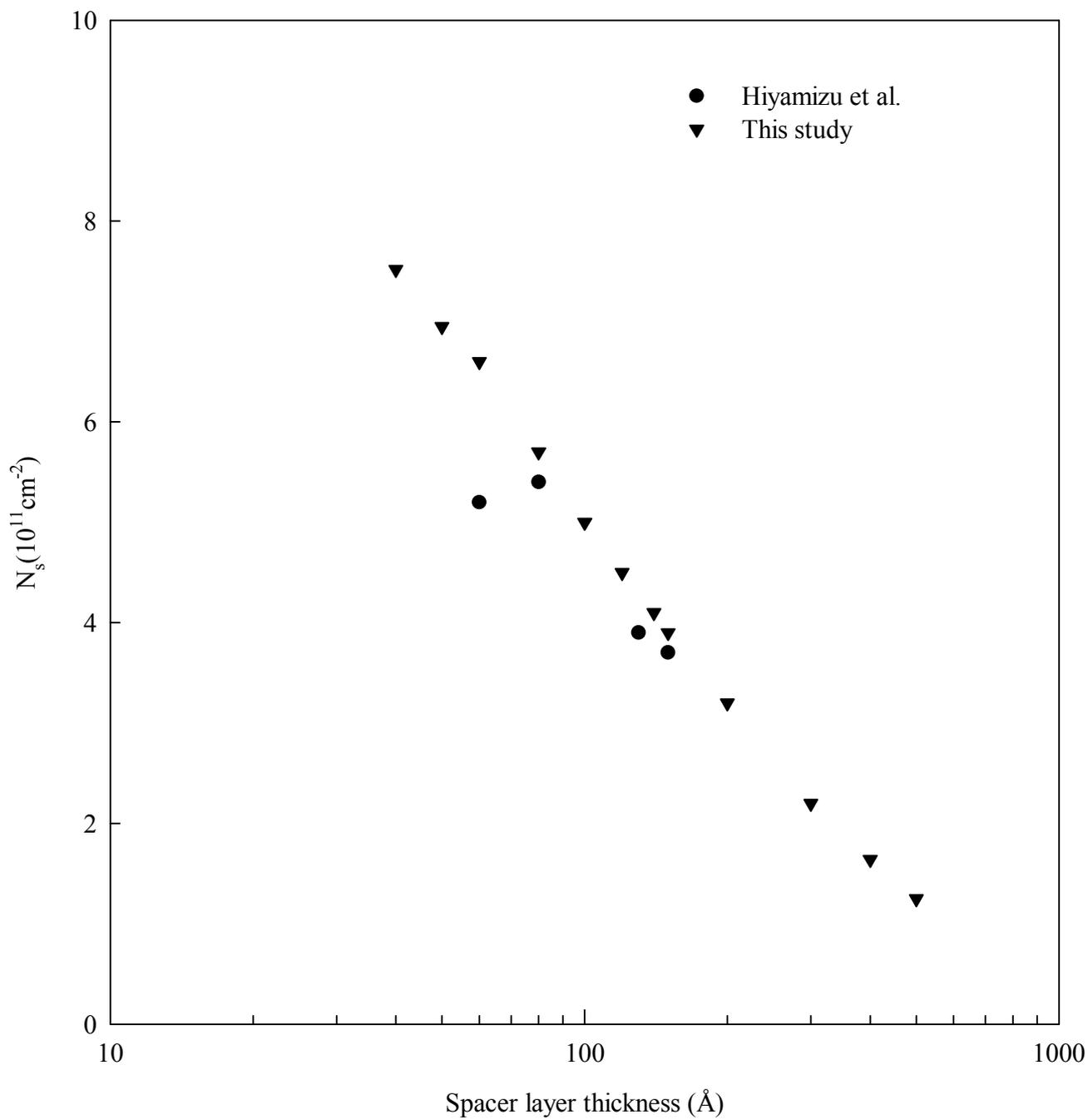

**Figure 4**

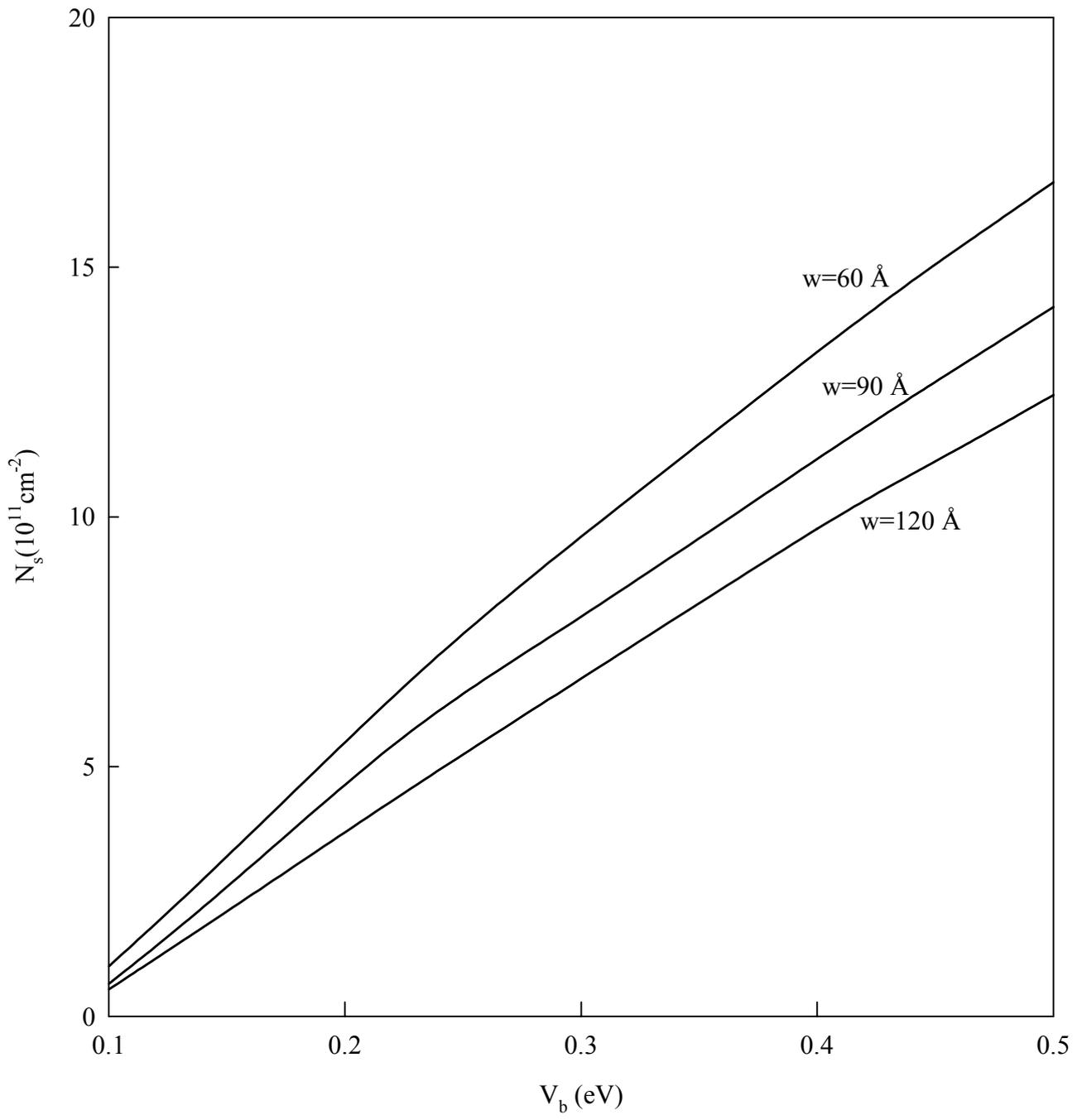

**Figure 5**